\documentclass[aps,prl,preprint]{revtex4-1}
\usepackage{graphicx}
\usepackage{amsfonts}
\usepackage{amsmath}
\usepackage{amssymb}
\usepackage{color}
\usepackage{bm}
\usepackage{amssymb}

\usepackage{multirow}
\usepackage{epstopdf}
\usepackage{natbib}

\newcommand{\beq}{\begin{equation}}
\newcommand{\eeq}{\end{equation}}

\def\R{\mathbf{\hat{R}}}

\def\jcp#1#2#3{{\it J.~Chem.~Phys.}~{\bf #1},\ #2\ (#3)}

\def\R{\mathbf{\hat{R}}}
\def\mathbi#1{\textbf{\em #1}}

\def\colvecnext#1{
        #1
        \global\advance\colveccount-1
        \ifnum\colveccount>0
                \\
                \expandafter\colvecnext
        \else
                \end{pmatrix}
        \fi
}

\newcommand{\mat}[2][ccccccccccccccccccccccccccccccccccccccccccccccccccccccccccccc]{\left(
                                                \begin{array}{#1}
                                                    #2\\
                                                \end{array}
                                                \right) }

\begin{document}

\title{
Efficient non-parametric fitting of potential energy surfaces for polyatomic molecules with Gaussian processes}
\author{Jie Cui}
\affiliation{Department of Chemistry, University of British Columbia, Vancouver, B.C., V6T 1Z1, Canada}

\author{Roman V. Krems}
\affiliation{Department of Chemistry, University of British Columbia, Vancouver, B.C., V6T 1Z1, Canada}
\date{\today}

\begin{abstract}
We explore the performance of a statistical learning technique based on Gaussian Process  (GP) regression as an efficient non-parametric method for constructing multi-dimensional potential energy surfaces (PES) for polyatomic molecules. Using an example of the molecule N$_4$, we show that a realistic GP model of the six-dimensional PES can be constructed with only 240 potential energy points. We construct a series of the GP models and illustrate the convergence of the accuracy of the resulting surfaces as a function of the number of {\it ab initio} points.  We show that the GP model  based on $\sim 1500$ potential energy points achieves the same level of accuracy as the conventional regression fits based on 16,421 points. The GP model of the PES requires no fitting of {\it ab initio} data with analytical functions and can be readily extended to surfaces of higher dimensions.

\end{abstract}

\maketitle

\section{Introduction}

Accurate potential energy surfaces (PES) describing the interactions of atoms in polyatomic molecules are required for calculations of the ro-vibrational energy levels and the dynamical properties of molecules. While classical dynamics calculations may use local values of the potential energy and its gradient, quantum dynamics calculations generally require the knowledge of the global PES. Obtaining the global PES for a polyatomic molecule requires quantum chemistry calculations of the potential energy at different molecular geometries and the construction of an analytical fit to interpolate the computed energies. The resulting PES must be continuous, smooth, differentiable and free of unphysical variations. As the complexity of the molecule increases, producing an analytical fit that satisfies these requirements becomes exceedingly difficult \cite{high-dimensional-PES}. The difficulty arises from (i) the need to calculate the potential energy at a large number of coordinates; (ii) the complexity of the analytical functions and fitting procedures necessary for representing multi-dimensional surfaces.

Several methods have been developed recently to reduce the difficulty of fitting PESs for polyatomic molecules. For example, an $n$-mode representation was proposed to construct a PES as a series of intrinsic potentials accounting for a small number of normal modes \cite{n-node-1,n-node-2,n-node-3}. An alternative approach -- or one that can be used in combination with the $n$-node representation -- is to take advantage of the symmetry of multi-dimensional PES under permutations of identical atoms \cite{high-dimensional-PES, high-D-PES2, high-D-PES3}. The goal of these approaches is to fit a high-dimensional PES by fitting functions of lower dimensionality. However, both of these approaches are complex and, therefore, difficult to implement; are molecule-specific (i.e. the fitting procedure and the choice of analytical functions are different for different molecular species and different number of active degrees of freedom); and meet with the same challenges as the conventional fitting procedures when the dimensionality of the PES increases. 

In order to construct accurate PES for large polyatomic molecules or molecular complexes, it is desirable to develop alternative approaches that would be 
\begin{itemize}

\item[(i)] easy to implement; 
\item[(ii)] scalable;
\item[(iii)] universal (be the same for any molecule and any number of dimensions); 
\item[(iv)] efficient (i.e. using a small number of {\it ab initio} calculations per dimension).

\end{itemize} 
This can be achieved by combining quantum chemistry calculations with machine-learning techniques developed for efficient interpolation of multi-dimensional spaces \cite{ml}. There are, at least, two machine-learning methods that can be used for obtaining high-dimensional PESs: an approach based on artificial neural networks \cite{nn-1} and  Gaussian process regression \cite{gp-model1}. 
The application of artificial neural networks to fitting PESs has been explored in several studies \cite{nn-2,nn-3,nn-4,nn-5,nn-6,nn-7,nn-8}. Gaussian process (GP) models have been proposed for constructing force-fields in Refs. \cite{gp-1,gp-2,gp-3,gp-4,gp-5} and for a variety of applications in molecular collision dynamics in Refs. \cite{jie-1,jie-2}.  However, to the best of our knowledge, the accuracy and efficiency of GP regression for obtaining a global PES for a polyatomic molecule have not been systematically assessed before. It is generally not known how many {\it ab initio} points are required for a GP regression to produce a physical PES and whether the error of GP fits can be reduced to the accuracy of {\it ab initio} calculations.

In the present work, we explore the efficiency of Gaussian processes for interpolating multi-dimensional PESs by producing a series of GP models 
of the six-dimensional PES for N$_4$. We use the potential energy data computed by Paukku et al.~\cite{n4} to train the GP models and explore the accuracy of the resulting surfaces as a function of the number of training points.
This PES covers a wide range of energies between $0$ and $1000$ kcal/mol and exhibits a complex energy landscape. 
 We show that a  realistic model representing accurately the low-energy part of the 6D PES can be obtained with only 240 quantum chemistry calculations. 
  We illustrate that the GP model based on between 1200 and 1800 potential energy points achieves the same level of accuracy as the conventional regression fits based on 16,421 points, capturing the features of the PES even at energies as high as 1000 kcal/mol. 

\section{Why Gaussian Processes for fitting PES?}

Given the abundance of literature on the application of the artificial neural networks to fitting PESs  \cite{nn-2,nn-3,nn-4,nn-5,nn-6,nn-7,nn-8}, why to explore an alternative statistical learning technique? 
Gaussian processes offer, at least, five major advantages for obtaining PESs: 

\begin{itemize}

\item[(i)] GP regression is a kernel-based statistical learning technique and, as such, is generally easier to implement than the artificial neural networks \cite{gp-1}. 

\item[(ii)] There is no need to fit any computed data with analytical functions \cite{gp-model1,gp-model2,gp-model3}. A GP model is determined by correlations between potential energy points in a multi-dimensional configuration space and provides a statistical prediction for the value of the potential energy within the configuration space. The correlations do not need to be known exactly; rather, one uses the best estimates of the correlations parametrized by a simple analytical function. 

\item[(iii)] With a proper parametrization of correlations between the energy points, a GP model is guaranteed to yield a smooth and differentiable surface that passes through the potential energy points used for training the model \cite{mitchell1990existence, cressie1993statistics, stein1999interpolation,abt1999estimating}. The choice of the correlation function controls the differentiability of the resulting surface. 

\item[(iv)] The GP models scale favorably with the dimensionality of the problem. A rule of thumb is that a GP model should require on the order of $10 \times q$ points for interpolating a surface with $q$ dimensions \cite{sample-size}. This implies
that a 6D PES can be obtained with only 60 quantum chemistry calculations. The purpose of this work is to test this rule of thumb in application to fitting the molecular PESs, typically characterized by smooth but wide variation of energy and the presence of multi-dimensional minima and 
barriers in the intramolecular coordinate space. Artificial neural networks are expected to require many more {\it ab initio} points \cite{nn-8}.

\item[(v)] GP models are guaranteed to become more accurate when trained by more quantum chemistry calculations \cite{sample-size}. 

\end{itemize}

As will be clear from the discussion in the following section, the numerical effort associated with training a GP model is $\mathcal{O}(n^3)$, 
where $n = (p \times q)^3$ is the number of {\it ab initio} training points, $q$ is the dimensionality of the PES and $p$ is the effective number of {\it ab initio} points per dimension. 
Once a GP model is trained, the evaluation of the potential energy using the GP model is reduced to a product of a square matrix and a column vector with the dimension equal to the number of training points. Thus, the numerical efficiency of the PES evaluation based on a GP model is 
$\mathcal{O}(n^2)$. For applications requiring the evaluation of the PES at a large number of intramolecular coordinates, 
this may become a bottleneck when  $p \times q \approx 10^3$ -- $10^4$. One should, therefore, expect the GP regression to be a method of preference for dimensions $q \lesssim 100$ and
the artificial neural network approach -- for problems with large $q \gtrsim 100$.

\section{Fitting PES with a Gaussian Process}

We will denote a GP by $F(\cdot)$.
A GP  is a family of normally distributed random multi-dimensional functions characterized by a mean function $\mu(\cdot)$ and a covariance function $K(\cdot, \cdot)$. For a GP with constant variance $\sigma^2$, the covariance function is $K(\cdot, \cdot) = \sigma^2 R(\cdot, \cdot)$, where $R(\cdot, \cdot)$ is a correlation function. A point in a multi-dimensional space of interest is specified by a vector $\bm x$. A realization of a GP at a given $\bm x$ is the value of a random function drawn from the normal distribution and evaluated at  $\bm x$. Multiple outputs of a GP at the same $\bm x$ produce a Gaussian distribution of values $F(\bm x)$.

The application of the GP models to molecular dynamics problems has been described in our previous work \cite{jie-1,jie-2}. Here, we assume that the internal coordinates of a polyatomic molecule are given by a $q$-dimensional vector $\bm x = ( x_1, ..., x_q )^\top$. Our goal is to construct the global PES, given $n$ potential energy values at vectors $\bm x_1, ..., \bm x_n$. We model the collection of $n$ points in the $q$-dimensional space by a GP and assume that each of these potential energies $V(\bm x_i)$ computed by a quantum chemistry method is a realization of a GP at $\bm x_i$.  
The multiple outputs of a GP at the given $n$ points  $\bm{Y}^n=\Big({F}(\bm{x}_1), 
{F}(\bm{x}_2),\cdots,{F}(\bm{x}_n)\Big)^{\top}$ follow a multivariate normal distribution
\begin{eqnarray}
\bm{Y}^n \thicksim \mathrm{MVN} (\boldsymbol{\beta},\sigma^2 \mathbf{A})
\end{eqnarray}
where $\boldsymbol{\beta}$ is the mean vector and $\mathbf{A}$ is a $n\times n$ matrix defined as
\begin{eqnarray}
\mathbf{A} =  \left(\begin{array}{cccc} 1 & R(\bm x_1, \bm x_2) & \cdots & R(\bm x_1, \bm x_n) \\  R(\bm x_2, \bm x_1) & 1& \ & \vdots \\ \vdots&  & \ddots &  \\R(\bm x_n, \bm x_1) & \cdots &  & 1 \\\end{array} \right)
\end{eqnarray}
For simplicity, and without loss of generality, we assume that $\boldsymbol{\beta} = \beta {\bf I}$, where $\bf I$ is the identity vector of dimension $n$ and $\beta$ is an unknown scalar parameter.

 We describe the correlation function $R(\cdot, \cdot)$  by the following expression \cite{mitchell1990existence, cressie1993statistics, stein1999interpolation,abt1999estimating}:
 \begin{eqnarray}
R(\mathbi{x},\mathbi{x}') = \left\{ \prod_{i=1}^{q} \bigg(1+\frac{\sqrt{5}|x_i-x_i'|}{\omega_i}+\frac{5(x_i-x_i')^2}{3\omega_i^2}\bigg)\mathrm{exp}\bigg(-\frac{\sqrt{5}|x_i-x_i'|}{\omega_i}\bigg) \right\},
\label{correlation}
\end{eqnarray}
where $\omega_i$ are the unknown parameters representing the characteristic length scales of the correlations.
Eq. (\ref{correlation}) is a special case of the Mat\'ern correlation function \cite{mitchell1990existence, cressie1993statistics, stein1999interpolation,abt1999estimating} defined as 
\begin{eqnarray}
R(\mathbi{x},\mathbi{x}') =\prod_{i=1}^{q}\frac{1}{\Gamma(\nu)2^{\nu-1}}d_i^{\nu}\mathcal{K}_{\nu}(d_i)
\label{correlation-matern}
\end{eqnarray}
 where $d_i = \sqrt{2\nu}|x_i - x_i'|/\omega_i$ and $\mathcal{K}_{\nu}(\cdot)$ is the modified Bessel function of order $\nu$. Eq.~(\ref{correlation}) represents the Mat\'ern correlation function with $\nu = 5/2$. 
The mathematical form of the correlation function determines the properties of the resulting GP. In particular, it determines the existence of its derivatives. 
For example, if the correlation function is chosen to be a Gaussian \cite{mitchell1990existence, cressie1993statistics, stein1999interpolation,abt1999estimating},  the GP is differentiable to any order. With the Mat\'ern correlation function (\ref{correlation-matern}), the process is differentiable to order $k<\nu$. Thus, when the parameter $\nu=5/2$, the GP is twice differentiable.

To find the parameters $\boldsymbol{\omega}=(\omega_1,\omega_2,\cdots,\omega_q)^\top$ of the correlation function making the predicted potential energy points ``most likely'', known in statistics as the 
maximum likelihood estimators (MLE), we maximize numerically the log-likelihood function
\begin{eqnarray}
\textrm{log}\mathcal{L}(\boldsymbol{\omega}|\bm{Y}^n) = -\frac{1}{2}\left[n\textrm{log}\hat{\sigma}^2+ \textrm{log}(\textrm{det}(\textbf{A})) +n \right],
\label{log-L}
\end{eqnarray}
where 
\begin{eqnarray}
\hat{\sigma}^2 (\boldsymbol{\omega}) = \frac{1}{n}(\textbf{\textit{Y}}^n-\boldsymbol{\beta})^{\top}{\bf{A}}^{-1}(\textbf{\textit{Y}}^n-\boldsymbol{\beta}),
\end{eqnarray} 
\begin{eqnarray}
{\hat{\beta}} (\boldsymbol{\omega}) = (\mathbf{I}^{\top}\textbf{A}^{-1}\mathbf{I})^{-1}\mathbf{I}^{\top}\textbf{A}^{-1}\textbf{\textit{Y}}^n,
\end{eqnarray}
and the hat over the symbol denotes the MLE. 

The goal is to make a prediction of the potential energy value at an arbitrary position $\bm x = \bm x_0$. 
The values $Y_0=F(\bm{x}_0)$ obtained by multiple realizations of the GP at $\bm x_0$
and the multiple outputs of the GP at training sites 
$\bm{Y}^n=\Big({F}(\bm{x}_1), {F}(\bm{x}_2),\cdots,{F}(\bm{x}_n)\Big)^{\top}$ 
are jointly distributed as
\begin{eqnarray}
\mat{Y_0\\ \bm{Y}^n}  \thicksim \mathrm{MVN} \left\{\mat{1 \\ \textbf{I}}{\beta}
, \sigma^2 \mat{1 & \mathbf{A}_0^{\top}\\ \mathbf{A}_0&\mathbf{A} }\right\}
\end{eqnarray}
where 
 $\mathbf{A}_0=(R(\bm{x}_0,\bm{x}_1),R(\bm{x}_0,\bm{x}_2),\cdots,R(\bm{x}_0,\bm{x}_n))^{\top}$ is a column vector specified by the correlation function $R(\cdot|\boldsymbol{\hat \omega})$ with the MLE of $\boldsymbol{\omega}$. This means that 
the conditional distribution of values $Y_0=F(\bm{x}_0)$ given the values $\bm{Y}^n$ is a normal distribution 
with the conditional mean 
\begin{eqnarray}
\label{mean-prediction}
\tilde \mu(\bm{x}_0)&=& {\beta}+\mathbf{A}_0^{\top}\mathbf{A}^{-1}(\bm{Y}^n -\boldsymbol{\beta} ) 
\end{eqnarray}
and the conditional variance
\begin{eqnarray}
\tilde \sigma^{2}(\bm{x}_0) &=& \sigma^2 (1- \mathbf{A}_0^{\top}\mathbf{A}^{-1}\mathbf{A}_0).
\label{conditional-sigma}
\end{eqnarray}
We use Eq. (\ref{mean-prediction}) as a prediction of the value of the potential energy at $\bm x_0$ and Eq. (\ref{conditional-sigma}) as the uncertainty of the prediction.

\section{Results}

Following Ref. \cite{n4}, we illustrate the quality of the PES model by Figures 1-2 showing the potential energy curves for the dissociation N$_4 \rightarrow$ N$_3$ + N at different values $d$ of the distance between the centers of mass of the two nitrogen molecules. The different panels of Figures 1-2 correspond to different geometries of the N$_4$ complex, covering the $A$-shaped, $T$-shaped, $H$-shaped and $X$-shaped sets illustrated in Figure 3 of Ref. \cite{n4}. 
The symbols in Figures 1-2 represent the original potential energy data and the curves -- the values computed by the GP models.

In principle, the potential energy points in vector $\bm Y^n$ used for training the GP model can be chosen at random configurations of the molecule $\bm x_i$. However, given a fixed number of training points, the accuracy and the stability of the GP model can be improved if the points are selected to cover evenly the configuration space of interest.  This can be achieved, for example, with the   
Latin hypercube sampling (LHS) method \cite{LHS} known to cover a multi-dimensional space efficiently. 
In order to improve the efficiency of the GP models, we use the following sampling technique for the results shown in Figures 1-2. Paukku et al. \cite{n4} computed 15363 points for 
9 combinations of internal angles (9 sets of geometries) for N$_4$ and 1056 points for 
77 combinations of internuclear distances for N$_3$. We first treat each combination of
three input variables as a single variable to get a reduced 4D PES.  We then generate an $n \times4$
LHS matrix with values uniformly distributed on $[0,1]$, and transform elements
in each row to sample quantiles corresponding to the empirical distribution of input variables in
the reduced 4D PES. If the sampled configuration is not contained in the data set of Paukku et al. \cite{n4}, we use the nearest available configuration. It would be better to use the LHS in six dimensions. However, that would lead to a lot of configuration points not contained in the data of Paukku et al. \cite{n4}. The sampling technique used here requires a much smaller number of adjustments to the sampled configurations.

Figure~1 is obtained with only 240 potential energy points. Given the small number of {\it ab initio} points, the agreement between the {\it ab initio} data (symbols) and the results of the GP model is remarkable. Figure 1 illustrates that 240 potential energy points produce a fitted GP model representing accurately the low-energy part ($< 200$ kcal/mol) of the global 6D PES and giving a qualitatively correct representation of the PES at high energies ($> 200$ kcal/mol).

The results of Figure 2 (obtained with 1200 potential energy points) show that increasing the number of potential energy points improves the quality of the model both at low energies and in the high energy region of the surface. 
In order to quantify the accuracy of the GP models of the PES and compare the result with those in Ref. \cite{n4}, we calculate the mean unsigned error (MUE) and the root-mean-square error (RMSE) 
for a series of GP models obtained with different numbers of potential energy points. 
The errors are defined as 
\begin{eqnarray}
{\rm MUE} = \frac{1}{n} \sum_{i = 1}^{n} |\hat{y}_i - y_i|,
\end{eqnarray}
\begin{eqnarray}
{\rm RMSE} = \sqrt{\frac{1}{n} \sum_{i = 1}^{n} |\hat{y}_i - y_i|^2},
\end{eqnarray}
where $\hat{y}_i$ is the value of energy obtained from the GP model and $y_i$ is the corresponding value of energy from the data of Paukku et al. \cite{n4}.
To ensure that the accuracy of the GP models shown in Figures 1 - 2 is not accidental, we construct up to 100 different GP models for each number of {\it ab initio} points summarized in Table I and compute the average errors with the corresponding standard deviations of the errors. 
The number of the GP models was chosen to ensure convergence of both the mean values and the standard deviations reported in Table I. 

Table I illustrates that the GP model trained by 1800 potential energy points produces the PES with the similar accuracy as the fitting procedure of Ref. \cite{n4} based on polynomial regressions and 16421 potential energy points. Table I also illustrates the monotonous improvement of  both the model accuracy and the model stability as the number of potential energy points increases. The accuracy of the GP model increases by almost 40 \% as the number of potential energy points increases from 240 to 480, and another 50 \% as the number of points increases from 480 to 1800. The increase of the stability is evident from the decrease of the standard deviations. 
The data in Table I illustrate that the low-energy part of the PES corresponding to $E< 100$ kcal/mol can be represented with the RMSE $< 2$ kcal/mol by a GP model with only 720 potential energy points.

\section{Conclusion}

We have constructed a series of Gaussian Process models of the global 6D potential energy surface for the molecule N$_4$ and showed that an accurate surface spanning a wide range of energies between $0$ and $1000$ kcal/mol can be obtained with only $\sim 1500$ potential energy points. The number of potential energy points required for an accurate GP model of the PES can be decreased by selecting the points to follow a 6D LHS \cite{LHS} instead of a quasi-LHS used here and by using regression functions to approximate the unconditional mean function $\mu(\cdot)$. For example, the mean of the GP can be modelled as 
\begin{eqnarray}
\mu(\bm x) = \sum_{j=1}^{s} {h}_{j}(\bm{x})\beta_{j} = {\bf{h}}(\bm{x})^\top{\boldsymbol{\beta}}
\label{mean}
\end{eqnarray}
where ${\bf h} = \left ( h_1(\bm x), ..., h_s(\bm x) \right )^\top$ is a vector of $s$ regression functions \cite{gp-model2,gp-model3} chosen to mimic the dependence of the PES on the corresponding variables. Since the variation of the PES with the individual bond lengths is often nearly quadratic at low energies and follows simple power laws in the limit of bond dissociations, it should often be possible to find suitable regression functions $h_{j}$ to optimize the GP model training. We note that the final results do not depend on the specific form of the functions $h_s(\bm x)$, as is clear from the calculations in the present article based on the simple choice of $h_1 = 1$ and $h_{s>1}=0$.

Our results have two important implications. First, a GP model, even with the simple choice of $h_1 = 1$ and $h_{s>1}=0$, requires much fewer potential energy calculations for the construction of an accurate global potential surface than conventional fitting techniques based on polynomial regressions. Second, the extension of the GP model discussed here to surfaces of higher dimensions is straightforward. 
Adding another dimension simply requires an extension of the expression for the correlation function  to include another term in the product (\ref{correlation}). Adding another dimension also requires more training points in the input vector $\bm Y^n$. However, it is known that the number of additional training points required per added dimension decreases with the number of dimensions \cite{sample-size}. The numerical procedure of training the GP model is limited by the optimization of the log-likelihood function (\ref{log-L}), involving iterative computations of the determinant and inverse of the matrix $\bf A$. The current limit on the number of dimensions amenable to GP modelling is limited to about 100 \cite{maximum-size} and there is currently active research on extending the applications of GP models to problems of higher dimensionality \cite{bigger-size}. It is thus foreseeable that the GP model discussed here can be applied for constructing the  PES for molecules with up to 100 internal degrees of freedom. Such a PES model would require an estimated 10,000 to 30,000 potential energy points, which is currently well within reach of state-of-the-art quantum chemistry calculations.


\section{Acknowledgment}

This work is supported by NSERC of Canada.

\clearpage
\newpage

\begin{table}[h!]
\centering
\caption{The mean unsigned errors (MUE in kcal/mol) and root-mean-square errors (RMSE in kcal/mol) of the 
Gaussian process models of the PES for N$_4$ trained by the different numbers of potential energy points. 
The errors of the Gaussian process models are obtained by averaging over up to 100 models with different samples of the potential energy points and reported with one unit of the corresponding standard deviations. The results in columns 3 and 4, labeled ``low $E$'', are for the potential energy $E < 100$ kcal/mol. The results in columns 5 and 6, labeled ``all $E$'', give the errors of the global PES over the entire range of energies extending to $E > 1000$ kcal/mol.}
\label{my-label}
\begin{tabular}{|c|c|c|c|c|c|}
\hline
\multicolumn{1}{|l|}{} & \multicolumn{1}{l|}{Number of points} & MUE (low $E$)&RMSE (low $E$)&MUE (all $E$) & RMSE (all $E$)\\ \hline
PES of Ref. \cite{n4} & 16421&1.3&1.8 & 5.0 & 14.3 \\ \hline
\multirow{5}{*}{GP model} & 240 &4.43 $\pm$ 1.37 & 6.39 $\pm$ 1.77 & 13.97 $\pm$ 2.02 & 38.00 $\pm$ 7.92 \\ \cline{2-6} 
                                        & 480 & 2.61 $\pm$ 0.99& 4.19 $\pm$ 1.56  & 9.68 $\pm$ 1.05  & 28.67 $\pm$ 3.92  \\ \cline{2-6} 
                                        & 720  &1.72 $\pm$ 0.52&2.96 $\pm$ 0.78  & 7.72 $\pm$ 0.78  & 24.02 $\pm$ 3.15  \\ \cline{2-6} 
                                        & 960  &1.39 $\pm$ 0.34&2.58 $\pm$ 0.67  & 6.57 $\pm$ 0.77 & 21.57 $\pm$ 2.78  \\ \cline{2-6} 
                                        & 1200  &1.23 $\pm$ 0.28&2.26 $\pm$ 0.49  & 5.76 $\pm$ 0.59  & 19.46 $\pm$ 2.75  \\ \cline{2-6} 
                                        & 1800   & 0.96 $\pm$ 0.23&1.81 $\pm$ 0.46  & 4.45 $\pm$ 0.57  & 15.87 $\pm$ 2.72  \\ \cline{2-6} 
                                        & 2400   & 0.74 $\pm$ 0.19& 1.50 $\pm$ 0.37 & 3.83 $\pm$ 0.53  & 14.33 $\pm$ 2.38  \\ \hline
\end{tabular}
\end{table}

\begin{figure}[ht]
\label{figure1}
\begin{center}
\includegraphics[scale=0.3]{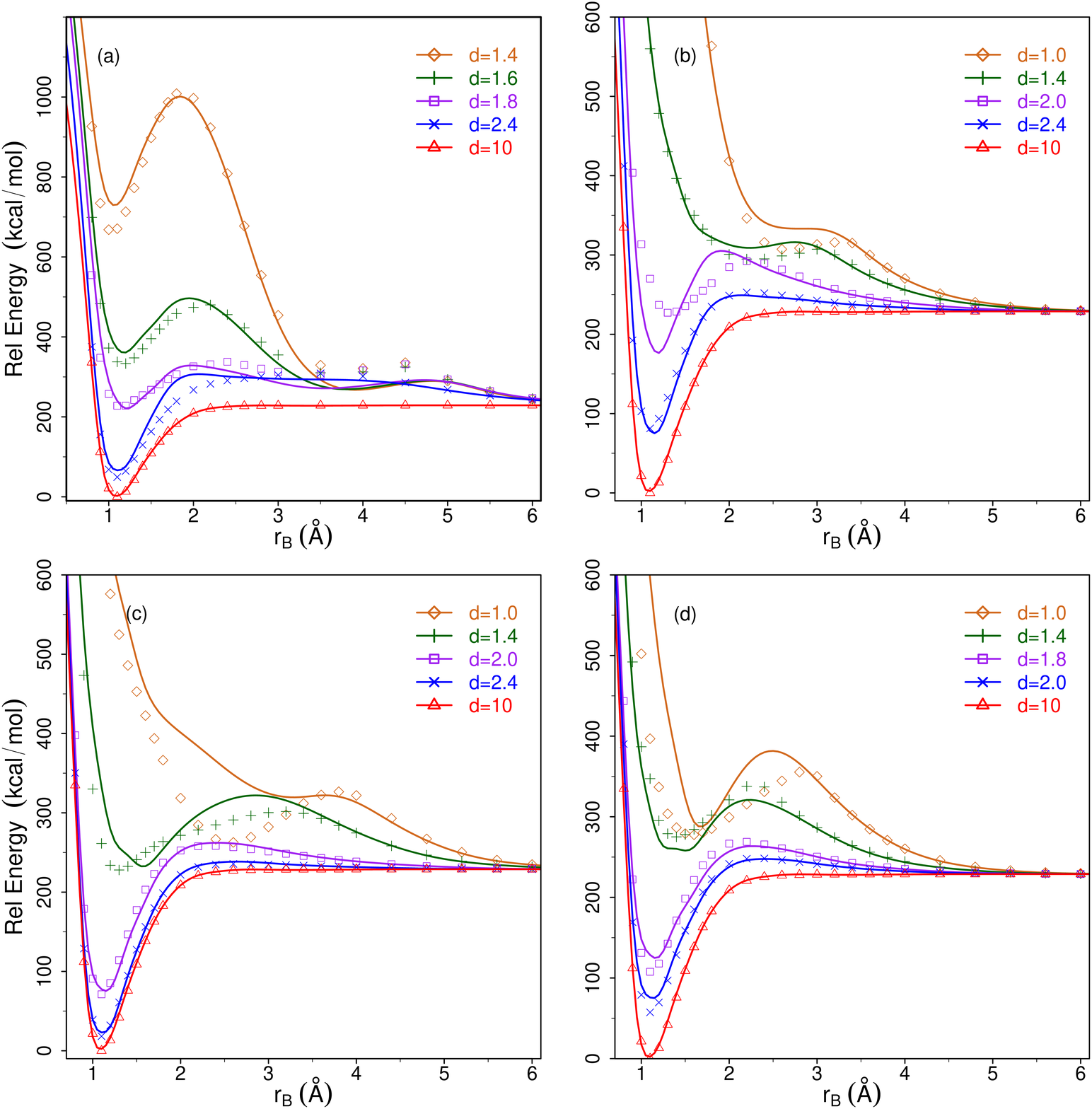}
\end{center}
\caption{PES for N$_4$ represented by the GP model trained with 240 potential energy points from Ref. \cite{n4}. The curves represent the GP model and the symbols -- the {\it ab initio} data. The variable $r_B$ is the interatomic distance in one of the N$_2$ molecules. The interatomic distance of the other molecule is fixed to the equilibrium distance. The different curves correspond to different values of the separation $d$ between the centers of mass of the two molecules. The different panels correspond to different geometries of the N$_4$ complex: (a) $A$-shaped; (b) $T$-shaped; (c) $H$-shaped; (d) $X$-shaped. These geometries are illustrated in Figure 3 of Ref. \cite{n4}. 
}
\end{figure}

\begin{figure}[ht]
\label{figure1}
\begin{center}
\includegraphics[scale=0.3]{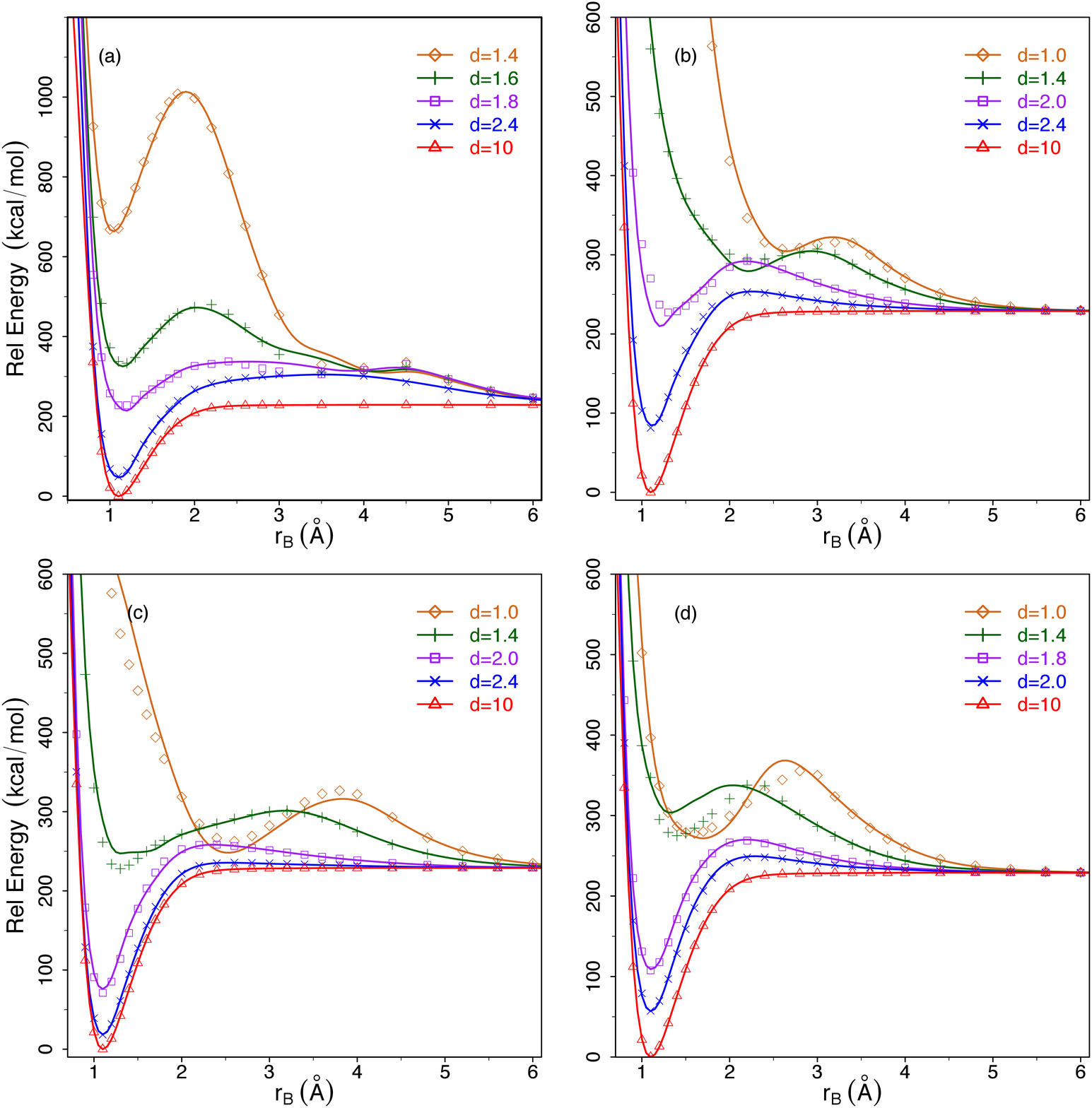}
\end{center}
\caption{Same as in Figure 1 but for the PES for N$_4$ represented by the GP model trained with 1200 potential energy points. 
}
\end{figure}



\begin{thebibliography}{99}






\bibitem{high-dimensional-PES}
B. J. Braams and J. M. Bowman, {\it Int. Rev. Phys. Chem.} {\bf 28}, 577 (2009). 


\bibitem{n-node-1}
 S. Carter,  S. J. Culik, and J. M. Bowman, {\it J. Chem. Phys.} {\bf 107}, 10458 (1997).

\bibitem{n-node-2}
K. Yagi, C. Oyanagi, T. Taketsugu, and K. J. Hirao, {\it J. Chem. Phys.}
{\bf 118}, 1653 (2003).

\bibitem{n-node-3}
G. Rauhut, V. Barone, and P. Schwerdtfeger, {\it J. Chem. Phys.} {\bf 125}, 054308 (2006).

\bibitem{high-D-PES2}
Y. Wang, B. J. Braams, J. M. Bowman, S. Carter, and D. P. Tew, 
{\it J. Chem. Phys.} {\bf 128}, 224314 (2008).

\bibitem{high-D-PES3}
R. Conte, C. Qu, and J. M. Bowman, {\it J. Chem. Theory Comput.} {\bf 11}, 1631 (2015).

\bibitem{ml}
M. Rupp, {\it Int. J. Quant. Chem.} {\bf 115}, 1058 (2015). 

\bibitem{nn-1}
K. Hornik, M. Stinchcombe, and H. White, {\it Neural Networks} 
{\bf 2}, 359 (1989).

\bibitem{gp-model1}
J. Sacks, S. B. Schiller, and W. J. Welch, {\it Technometrics} {\bf 31}, 41 (1989).

\bibitem{nn-2}
L. M. Raff, M. Malshe, M. Hagan, D. I. Doughan, M. G. Rockley, J. Komanduri,  {\it J. Chem. Phys.} {\bf 122}, 084104 (2005).

\bibitem{nn-3}
S. Manzhos, X. Wang, R. Dawes, and T. Carrington,  {\it J. Phys. Chem.}
A {\bf 110}, 5295 (2006).

\bibitem{nn-4}
C. M. Handley and P. L. A.  Popelier, {\it J. Phys. Chem.} A {\bf 114},
3371 (2010).

\bibitem{nn-5}
J. Behler and M. Parrinello,  {\it Phys. Rev. Lett.} {\bf 98}, 146401 (2007).

\bibitem{nn-6}
M. G. Darley, C. M. Handley, and P. L. A. Popelier, {\it J. Chem. Theory
and Comp.} {\bf 4},  1435 (2008).

\bibitem{nn-7}
J. Behler, {\it Phys. Chem. Chem. Phys.} {\bf 13}, 17930 (2011).

\bibitem{nn-8}
J. Chen, X. Xu, X. Xu, and D. H. Zhang, {\it J. Chem. Phys.} {\bf 138}, 221104 (2013). 


\bibitem{gp-1}
C. M. Handley, G. I. Hawe, D. B. Kellab
and P. L. A. Popelier,
{\it Phys. Chem. Chem. Phys.} {\bf 11}, 6365 (2009).

\bibitem{gp-2}
A. P. Bart\'ok, M. C. Payne, R. Kondor, and G. Cs\'anyi, {\it Phys. Rev. Lett.} {\bf 104}, 136403 (2010). 

\bibitem{gp-3}
A. P. Bartok and G. Csanyi, {\it Int. J. Quant. Chem.} {\bf 115}, 1051 (2015).

\bibitem{gp-4}
M. Caccin, Z. Li, J. R. Kermode, and A. De Vita, {\it Int. J. Quant. Chem.} {\bf 115}, 1129 (2015).

\bibitem{gp-5}
P. L. A. Popelier, {\it Int. J. Quant. Chem.} {\bf 115}, 1005 (2015).

\bibitem{jie-1}
J. Cui and R. V. Krems, {\it Phys. Rev. Lett.} {\bf 115}, 073202 (2015).

\bibitem{jie-2}
 J. Cui, Z. Li and R. V. Krems, {\it J. Chem. Phys.} {\bf 143}, 154101 (2015).


\bibitem{n4}
Y. Paukku, K. R. Yang, Z. Varga, and D. G. Truhlar, \jcp{139}{044309}{2013}.






\bibitem{gp-model2}
T. J. Santner, B. J. Williams, and W. I. Notz, {\it The Design and Analysis of Computer Experiments} (Springer Science $\&$ Bussiness Media, New York, 2003).

\bibitem{gp-model3}
C. E. Rasmussen and C. K. I. Williams, {\it Gaussian Process for Machine Learning} (The MIT Press, Cambridge, 2006).


\bibitem{mitchell1990existence}
T. Mitchell, M. Morris, and D. Ylvisaker, {\it Stoch. Proc. Appl.} {\bf 35}, 109 (1990).

\bibitem{cressie1993statistics}
N. A. C. Cressie, {\it Statistics for Spatial Data}, (Wiley-Interscience, New York, 1993).

\bibitem{stein1999interpolation}
M. L. Stein, {\it Interpolation of Spatial Data: Some Theory for Kriging}, (Springer Science $\&$ Business Media, 1999).

\bibitem{abt1999estimating}
M. Abt, {\it Scand. J. Stat.} {\bf 26}, 563 (1999).

\bibitem{sample-size}
J. L. Loeppky, J. Sacks, and W. J. Welch, {\it Technometrics} {\bf 51}, 366 (2009).


\bibitem{LHS}
M.~Stein,
{\em Technometrics} {\bf 29}, 143 (1987).


\bibitem{maximum-size}
V. Dubourg, B. Sudret, and F. Deheeger, {\it Probabilist. Eng. Mech.} {\bf 33}, 47 (2013).

\bibitem{bigger-size}
C. Linkletter, D. Bingham, N. Hengartner, D. Higdon and K. Q. Ye, {\it Technometrics} {\bf 48}, 478 (2006). 


\end{thebibliography}
\end{document}